\documentstyle[sprocl]{article}

\input{psfig}

\def\Journal#1#2#3#4{{#1} {\bf #2}, #3 (#4)}

\def\NPB{{\it Nucl. Phys.} B}
\def\PLB{{\it Phys. Lett.}  B}
\def\PRL{\it Phys. Rev. Lett.}

\def\ra{\rightarrow}

\def\be{\begin{equation}}
\def\ee{\end{equation}}
\def\bea{\begin{eqnarray}}
\def\eea{\end{eqnarray}}

\begin{document}

\title{SUB-MILLIMETER TESTS OF THE GRAVITATIONAL INVERSE SQUARE LAW} 

\author{E. G. ADELBERGER for the E\"OT-WASH GROUP}

\address{Center for Experimental Nuclear Physics and Astrophysics\\Box 354290, University of 
Washington\\
Seattle, WA 98195-4290, USA\\E-mail: eric@npl.washington.edu} 


\maketitle\abstracts{Sub-mm tests of the gravitational inverse-square law are 
interesting from several quite different perspectives. This paper discusses
work by the E\"ot-Wash group performed since the publication 
of our initial result in February 2001. We find no evidence for
short-range Yukawa interactions. Our results provide an upper limit
of 200~$\mu$m on the size of the largest ``extra'' dimension, and 
for the unification scenario with 2 large extra dimensions, set an upper
limit of 150~$\mu$m on the size of those dimensions.}

\section{What might be special about gravity at length scales below 1 mm?}
Very little is known about gravity at length scales below a 
few mm~\cite{lo:99}.
Recently theorists, using several different arguments, 
have suggested that the unexplored short-range regime of gravitation may hold 
profound surprises~\cite{ar:98,ar:00,di:96,su:97}, {\em i.e.}
that the gravitational interaction could display fundamentally new behavior
in the mm regime. 

Many of these arguments are based on the notion, 
inherent in string or M
theory, of more than 3 spatial dimensions. To 
maintain consistency with a vast body of observations the extra 
dimensions must be ``curled up'' in very small regions, usually
assumed to be comparable to 
$R_{\rm P}=\sqrt{G \hbar /c^3}=1.6 \times10^{-33}$~cm, 
or else hidden in some other way~\cite{ra:99}.
It has recently
been noted~\cite{ar:98,ar:00} that the enormous
discrepancy between natural mass scales of the Standard Model of
particle physics ($M_{\rm {SM}}\approx 1$ TeV) 
and of gravity (the Planck mass 
$M_{\rm P}=\sqrt{\hbar c/G}=1.2 \times 10^{16}$~TeV) 
could be eliminated if gravity propagates 
in {\em all} the space dimensions
while the other fundamental interactions are constrained to 
the three familiar dimensions. This unification scenario requires that
some of the
extra dimensions have radii $R^{\ast}$ that are 
large compared to $R_{\rm P}$ with
\begin{equation}
R^{\ast}=\frac{\hbar c}{M^{\ast} c^2} \left(  \frac{M_{\rm P}}{M^{\ast}}
\right)^{2/n}~,
\label{eq: R star}
\end{equation}
where $M^{\ast}$ is the unification scale (usually taken as $M_{\rm {SM}}$)
and $n$ is the number of large extra dimensions.
The scenario with $n=1$
is ruled out by astronomical data.
If there are 2 large extra dimensions, $R^{\ast}$  must be about 1 mm,
and the gravitational inverse-square law (which follows from
Gauss's Law in 3 spatial dimensions) will turn into a $1/r^4$-law
(Gauss's Law in 5 dimensions)
at distances much smaller than $R^{\ast}$.

Completely independent 
theoretical considerations also suggest that new effects may 
appear
at short distances; string theories predict scalar particles 
(dilatons and moduli) that generate Yukawa
interactions which could be seen in tests of the $1/r^2$
law. If supersymmetry is broken at low energies these scalar particles would
produce mm-scale effects~\cite{di:96,ka:00}.
Finally, there may be some significance to the observation~\cite{su:97}
that the gravitational 
cosmological constant, $\Lambda \approx 3$~keV/cm$^3$, deduced from distant
Type 1A supernovae~\cite{ri:98,pe:98} corresponds to a length scale 
$\sqrt[4]{\hbar c/\Lambda} \approx 0.1$~mm.
These, and other, considerations suggest that the Newtonian
gravitational potential should be replaced by a more general
expression~\cite{ke:99} 
\begin{equation}
V(r)=-G \frac {m_1 m_2}{r}(1 + \alpha e^{-r/\lambda})~.
\label{eq: potential}
\end{equation}
The simplest scenario with 2 large extra 
dimensions predicts $\lambda = R^{\ast}$ and 
$\alpha=3$ or $\alpha=4$ for compactification on
an 2-sphere or 2-torus, respectively~\cite{ke:99}, while dilaton and moduli
exchange could produce forces~\cite{di:96} with $\alpha$ as large as $10^5$ for
Yukawa ranges $\lambda \sim 0.1$~mm. 

\section{Experimental Results}
In February 2001 we published results of an inverse-square law 
test~\cite{ho:01} obtained with a novel torsion pendulum/rotating attractor 
instrument. The active component of the pendulum was an aluminum 
ring with 10 
equally-spaced holes bored into it. The pendulum was suspended just
above a disk-shaped copper attractor that had 10 similar holes 
bored into it. 
As the attractor rotated slowly and uniformly underneath the
pendulum, it produced a torque on the pendulum that varied back
and forth 10 times for every revolution of the attractor. The 
attractor actually consisted to two concentric disks each with
10 holes: a thinner upper disk and a thicker lower disk. The holes
in the lower disk were rotated by 18 degrees with respect to
those in the upper disk so that, if inverse-square-law were
correct, the torque on the
ring from the lower disk canceled the torque from the upper disk.
However, the torque from a short-range interaction could not be
canceled simply because the lower disk was too far away to
produce a short-range torque on the pendulum. We greatly reduced
any electrostatic torques on the pendulum by placing a stationary,
tightly stretched 20 $\mu$m thick Be/Cu membrane between the pendulum
and attractor. Our design had several attractive features:
\begin{enumerate}
\item the signal occured at a different frequency than the
disturbance (the revolution of the attractor). In this case
the signals were at $10\omega$, $20\omega$, and $30\omega$ where
$\omega$ is the attractor rotation frequency.  
\item our test bodies were the
``missing masses'' of the holes in cylindrical rings and disks.
This gave us accurately positioned test bodies with planar geometry 
(optimum because it maximizes the mass that can be placed in close 
proximity) that could be characterized very precisely.
\item the lower attractor disk that essentially canceled the 
Newtonian torque greatly reduced our sensitivity to nonlinearities
and scale-factor uncertainties in our instrument.
\end{enumerate}

\begin{figure}[t]
\hfil
\psfig{figure=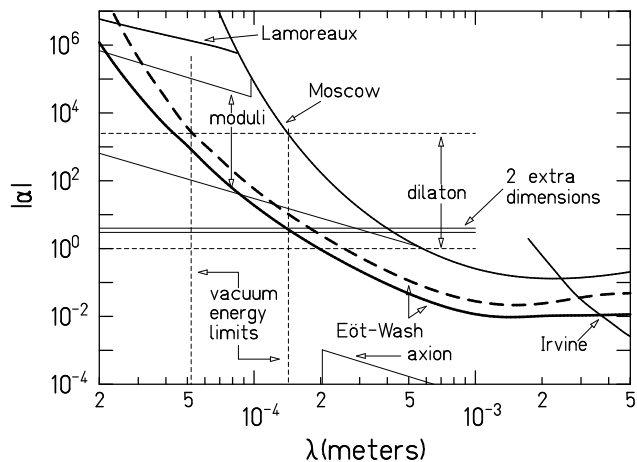,height=2.75in}
\caption{95\% confidence limits on a short-range Yukawa 
interaction. The heavy dashed line is the Mark II result from
Ref.~\protect\cite{ho:01}, the heavy solid line shows our new analysis 
of the combined Mark II and Mark III data.  The light 
lines showing some recent predictions are taken from
Ref.~\protect\cite{lo:99}.
\label{fig: constraints}}
\end{figure}

This experiment, which constituted the PhD thesis work of
C.D. Hoyle~\cite{ho:01a}, is described in Ref.~\cite{ho:01}.
The constraint on short-range Yukawa interactions from Ref.~\cite{ho:01a} is shown
in Fig.~\ref{fig: constraints}.
We encountered a surprising problem in the course of this measurement;
for a while looked as if we were observing a substantial violation
of the inverse-square law. Despite much effort, we could not account
for the apparent violation. So we constructed a second 10-hole torsion pendulum
and attractor having holes with different diameters and thicknesses to
check the original result and again saw an apparent violation of the 
$1/r^2$ law. Blayne Heckel finally identified the problem: the commercial
computer-controlled micropositioning stage from which the torsion fiber was
suspended had a scale factor error--it actually moved only $\approx 98\%$ 
as far as it indicated. So of course we did not find that 
$\vec{\nabla} \cdot \vec{g}=0$; we were using correct distances along
$\hat{x}$ and $\hat{y}$ and in incorrect distance along $\hat{z}$!
Reference~\cite{ho:01} was based on the results from the first (Mark II) 
instrument. Hoyle has recently reanalyzed the data from the Mark II instrument
as well as that from the second (Mark III) instrument. 
Figure~\ref{fig: constraints} shows the improved constraint from the
new analysis of the combined data. 
\section{Our second-generation instrument}
Since the publication of our original results~\cite{ho:01}, we have made
several improvements to our instrument. These were motivated by the
recognition that the torque from a very 
short-range Yukawa interaction with $\lambda \ll s$ ($s$ is the
closest attainable pendulum-to-attractor separation) scales as 
\begin{equation}
T=\frac{\Delta E}{\Delta \theta} \propto \rho_p \rho_a A \lambda^4 
e^{-(s/\lambda)}~,
\end{equation}
where $E$ is the interaction energy between the pendulum and attractor,
$\theta$ is the twist of the pendulum,
$\rho_p$ and $\rho_a$ are the densities of the pendulum and attractor, and
$A$ is the area of the holes.
\begin{enumerate}
\item We increased the torque from a given short-range Yukawa interaction
and reduced the Newtonian torque appreciably. This was done by:
\begin{itemize}
\item a new pendulum/attractor design that has two rows of 22 holes 
(to increase $A$) with
thinner (1 mm thick) pendulum ring and upper attractor disk. The relative
sizes of the penduolum and attractor holes was ``tweaked'' to put 
most of the power of a Yukawa torque into the fundamental $22 \omega$
signal.
\item the pendulum and attractor are both made from molybdenum; this 
increases the $\rho_p \rho_a$ product by a factor of 4.5.
\end{itemize}
\item The minimum attainable spacing $s$ is less by at least a factor
of two:
\begin{itemize}
\item we reduced the thickness of the conducting membrane to
$10~\mu$m.
\item we installed a passive damper that reduced the mean amplitude
of the pendulum bounce mode by a factor of 6. 
\end{itemize}
\item We cancelled Newtonian gravity to a higher degree by having
thinner active components with more and smaller holes (Newtonian
gravity, being long range, tends to average over several holes).
\item We reduced our torque noise by a factor of 6 by improving
the autocollimator performance and increasing the pendulum's 
quality factor $Q$ to $\approx 4000$.
\end{enumerate}

\begin{figure}[t]
\hfil
\psfig{figure=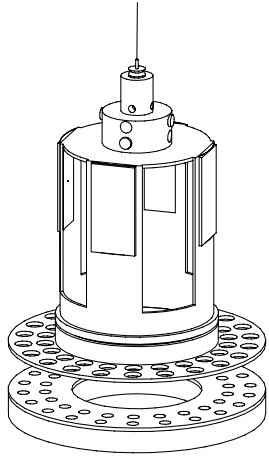,height=6cm}
\caption{Schematic view of the 22-fold rotationally 
symmetric Mark V instrument.
The active ring of the pendulum is molybdenum and has 44 holes.
The two attractor disks are also made from molybdenum. The upper 
disk has 44 holes while the lower, canceling disk has 22 larger 
holes situated between the holes in the upper disk. By making the attractor 
holes have a different diameter than those in the pendulum, we placed more
of the Yukawa signal into the fundamental $22\omega$ mode. 
\label{fig: 44-hole pendulum}}
\end{figure}

The pendulum/attractor of our current Mark V instrument is shown 
in Fig.~\ref{fig: 44-hole pendulum}.
\section{Conclusions}
We have tested the gravitational inverse-square law at length scales well
below 1 mm. We find no evidence for violations. For the proposed scenario
with 2 ``large'' extra dimensions, our negative result corresponds to
a 95\% confidence upper limit of 150 $\mu$m on the size of the 2 dimensions;
this implies a unification mass $M^{\ast}$ greater than 4.0 TeV.
Alternatively, we can use our results to set an upper limit on the size of
the {\em largest} extra dimension, regardless of how many there are (subject
only to the assumption that one of the extra dimensions is appreciably larger 
than the rest). In this case $\alpha=1$~\cite{di:01}, 
so that our data imply with
95\% confidence that $\lambda < 200$~$\mu$m. This upper limit 
constrains models that try to explain phenomena such as neutrino
oscillations in terms of large extra dimensions.
 
We expect that, in the next year or so, 
our torsion-balance scheme for testing the gravitational $1/r^2$ law 
will provide good results for length scales down to 50 $\mu$m. This is less
than the diameter of a typical human hair! If we do find evidence for
violation of the $1/r^2$ law, we would then develop an instrument to
check if the $1/r^2$ violating interaction also violates the
weak equivalence principle. This would distinguish between 
exotic space-time scenarios of extra-dimensions etc. that do not
violate the equivalence principle, from exotic particle exchange scenarios
that must violate the principle. 

However, until evidence for new physics
is found, it is clearly better to work on tests of the inverse-square law than 
on equivalence-principle tests: the $1/r^2$ tests are more general (probing
all finite-range effects), and more sensitive (in particle-exchange scenarios
the composition-dependence is expected to be a relatively small fractional
effect).  
But testing the gravitational $1/r^2$ law for length scales less than 50 
$\mu$m will 
probably require a somewhat different technology. In a planar geometry 
(optimum because one gets the maximum amount of
mass in close proximity) the signal
of a short-range Yukawa interaction drops as roughly the 4th power of
the Yukawa range while extraneous disturbances stay roughly the same
size. This will present an interesting challenge for future experimental
work.
\section*{Acknowledgments}%
Jens Gundlach, Blayne Heckel, CD Hoyle, Dan Kapner, Ulrich Schmidt and
Erik Swanson all made substantial contributions to the work reported here.
We are grateful to Profs. David Kaplan and Keith Dienes for illuminating
conversations. 
This work was supported by the NSF (grant PHY 9970987), and by DOE 
funding of the 
Center for Experimental Nuclear Physics and Astrophysics at the University
of Washington.

\end{document}